\documentclass[conference]{IEEEtran}
\IEEEoverridecommandlockouts

\usepackage{amsmath,amssymb,amsfonts}
\usepackage{algorithmic}
\usepackage{graphicx}
\usepackage{textcomp}
\usepackage{xcolor}
\usepackage[hidelinks]{hyperref}
\usepackage{svg}
\usepackage{multirow}
\usepackage{makecell}
\usepackage{booktabs}
\usepackage[acronym]{glossaries}
\usepackage{pifont}
\usepackage[maxbibnames=4,doi=false,style=ieee,backend=biber]{biblatex}
\usepackage[capitalise]{cleveref}
\def\BibTeX{{\rm B\kern-.05em{\sc i\kern-.025em b}\kern-.08em
    T\kern-.1667em\lower.7ex\hbox{E}\kern-.125emX}}

\begin{document}
\newacronym{vc}{VC}{Voice Cloning}
\newacronym{tts}{TTS}{Text-to-Speech}
\newacronym{for}{FoR}{Fake or Real}
\newacronym{gans}{GANs}{Generative Adversarial Networks}
\newacronym{mfccs}{MFCCs}{Mel-Frequency Cepstrum Coefficients}
\newacronym{lfccs}{LFCCs}{Linear Frequency Cepstral Coefficients}
\newacronym{gmm}{GMM}{Gaussian Mixture Model}
\newacronym{stft}{STFT}{Short Time Fourier Transform}
\newacronym{eer}{EER}{Equal Error Rate}

\newcommand{\etal}{\textit{et al}.\@ }
\newcommand{\ie}{\textit{i.e.},\@ }
\newcommand{\eg}{\textit{e.g.},\@ }

\title{DiffSSD: A Diffusion-Based Dataset \\For Speech Forensics\\
}

\newcommand{\td}{$^\dagger$}
\newcommand{\tdd}{$^\ddagger$}
\author{
\parbox{0.95\linewidth}{
    \hspace*{\fill} Kratika Bhagtani\td, Amit Kumar Singh Yadav\td, Paolo Bestagini\tdd, and Edward J. Delp\td \hspace*{\fill}\\
    \vspace*{0.1em}\\
    \small\centering \td School of Electrical and Computer Engineering, Purdue University, West Lafayette, Indiana, USA\\
    \small\centering \tdd Dipartimento di Elettronica, Informazione e Bioingegneria, Politecnico di Milano, Milano, Italy
}
}

\maketitle

\begin{abstract}
Diffusion-based speech generators are ubiquitous. 
These methods can generate very high quality synthetic speech and several recent incidents report their malicious use.
To counter such misuse, synthetic speech detectors have been developed.
Many of these detectors are trained on datasets which do not include diffusion-based synthesizers.
In this paper, we demonstrate that existing detectors trained on one such dataset, ASVspoof2019, do not perform well in detecting synthetic speech from recent diffusion-based synthesizers.
We propose the Diffusion-Based Synthetic Speech Dataset (DiffSSD), a dataset consisting of about 200 hours of labeled speech, including synthetic speech generated by 8 diffusion-based open-source and 2 commercial generators.
We also examine the performance of existing synthetic speech detectors on DiffSSD in both closed-set and open-set scenarios.
The results highlight the importance of this dataset in detecting synthetic speech generated from recent open-source and commercial speech generators.

\end{abstract}

\begin{IEEEkeywords}
Speech forensics, diffusion models, synthetic speech detection, speech dataset
\end{IEEEkeywords}

\glsresetall
\section{Introduction}\label{sec:intro}
\begin{figure}[ht]
\begin{center}
    \includegraphics[width= 0.8\columnwidth]{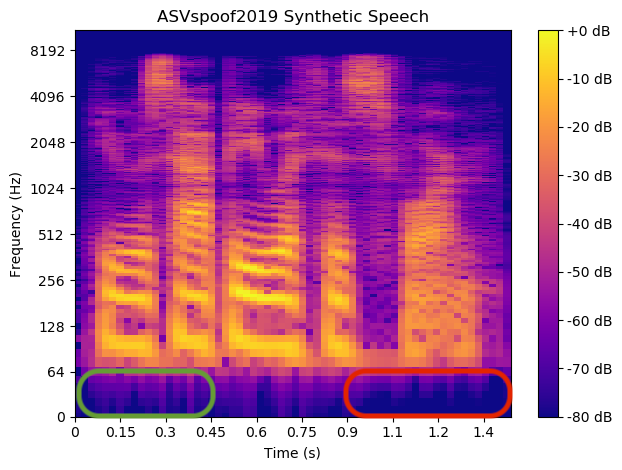}
\includegraphics[width= 0.8\columnwidth]{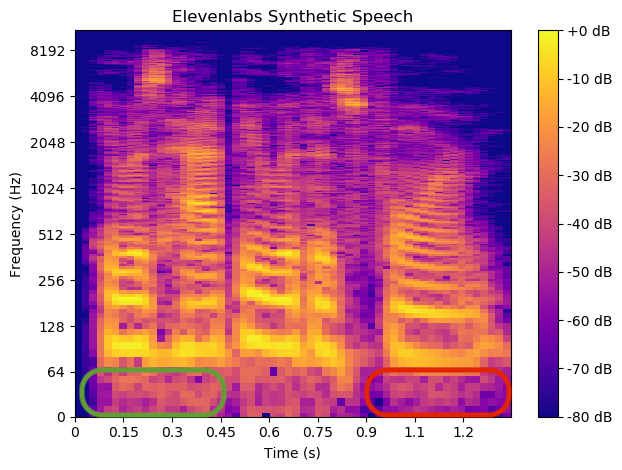}
\caption{Green and red boxes highlight some of the differences between spectrograms obtained for synthetic speech from a conventional speech generator~\cite{asvdata_2019} (top) and a recent commercial generator~\cite{elevenlabs2023} (bottom). Both signals have same speech content and speaker identity.
}
\label{fig:spectrogram}
\end{center}
\vspace{-3em}
\end{figure}
\begin{table*}
    \centering
    \caption{Common Datasets proposed for synthetic speech detection and their comparison with DiffSSD.}
    \resizebox{1.0\textwidth}{!}{
        {\renewcommand{\arraystretch}{1.1} 
        \begin{tabular}{lrrrrccc}
	        \toprule
            \textbf{Dataset} & \textbf{Speech} & \textbf{Real} &\textbf{Synthetic} & \textbf{Duration} & \textbf{Average Duration} &\textbf{Diffusion-based} & \textbf{Commercial}\\
            \textbf{Name} & \textbf{Synthesizers} & \textbf{Speech} & \textbf{Speech} & \textbf{(hours)} & \textbf{per speech (seconds)} &\textbf{Synthesizers} & \textbf{Synthesizers}\\
            \midrule
            ASVspoof2019 & 17 & 12,483 & 108,978  & 111.15 & 3.27 & \ding{55} & \ding{55} \\
            ASVspoof2021 & 17 & 16,492 & 148,148 & 132.51 & 2.63 & \ding{55} & \ding{55} \\
            FoR  & 7 & 111,000 & 87,285 & 149.67 & 3.17 &\ding{55} & \checkmark \\
            In-the-Wild & - & 19,963 & 11,816  & 37.85 & 4.29 & - & -\\
            TIMIT-TTS & 12 & 0 & 79,120 & 84.94 & 3.86 & \ding{55} & \checkmark \\
            \midrule
            \textbf{DiffSSD}  & \textbf{10} & \textbf{24,226} & \textbf{70,000} & \textbf{196.04} & \textbf{7.49} & \checkmark & \checkmark \\
	    \bottomrule
     \vspace{-2em}
        \end{tabular}}}
    \label{tab:dataset-comparison}
\end{table*}
\gls{vc} methods~\cite{arik2018,singh2024} are used to generate an individual's speech by mimicking the characteristics of their real voice. 
\gls{tts} \gls{vc} methods generate synthetic speech with spoken words corresponding to an input text~\cite{arik2018, bhagtani2024}.
While synthetic speech has useful applications in entertainment~\cite{films2022} and education~\cite{dai2022}, it is misused for fraud~\cite{smith2021}, misinformation~\cite{zelensky2022}, and malicious impersonation~\cite{david2018}. 

Speech Forensics focuses on authentication of speech to prevent misuse of synthetic speech~\cite{bhagtani2022overviewrecentworkmedia}.
Methods for detection and attribution of synthetic speech have been proposed~\cite{bhagtani2022overviewrecentworkmedia,tssdnet_2021,rahman2023attribution}.
For training and evaluating these methods, several datasets consisting of real and synthetic speech have been developed~\cite{yaroshchuk2023,zhang2022,zhang2022fmfcc}, \eg ASVspoof2019~\cite{asvdata_2019}, ASVspoof2021~\cite{asvspoof_2021}, \gls{for}~\cite{reimao2019}, In-the-Wild~\cite{müller2024} and TIMIT-TTS~\cite{salvi2023}.
Most~\gls{tts} generators in these datasets are conventional, \ie they use Recurrent Neural Networks (RNNs)~\cite{oord2016}, Hidden Markov Models (HMMs)~\cite{kayte2015hmm}, transformers~\cite{shen2018}, or ~\gls{gans}~\cite{salvi2023} for speech synthesis.
Table~\ref{tab:dataset-comparison} provides a summary of these datasets.
Some existing synthetic speech detectors~\cite{tssdnet_2021,tak22_odyssey,bartusiak2023,alzantot2019} have high detection performance on synthetic speech generated from these conventional generators.

Recently, more sophisticated speech generators have been proposed which use diffusion models~\cite{elevenlabs2023,huang2022prodiff,Liu2022DiffGANTTSHA,playht2023,kim2023_unitspeech}. 
These methods and commercial tools make high-quality voice cloning even more accessible for fraud.
Recent incidents have been reported about their misuse~\cite{ben2024},
making it essential to develop detectors that can detect synthetic speech generated from these tools.
For a fixed speaker identity and content, synthetic speech generated from such recent methods and conventional speech generators have differences as shown through a spectrogram in Figure~\ref{fig:spectrogram}. 
Noticing these differences, we conducted an experiment and concluded that it is not obvious for synthetic speech detectors trained on synthetic speech from conventional generators to  generalize and detect in these recent scenarios.
This raises the need for a comprehensive dataset with synthetic speech generated from recent diffusion-based methods and commercial tools.

In this paper, we propose Diffusion-Based Synthetic Speech 
Dataset (DiffSSD), 
consisting of about 200 hours of labeled speech, including
synthetic speech generated by 8 diffusion-based open-source and
2 commercial generators.
Overall, the dataset contains 70,000 synthetic speech signals from 11 distinct speakers.
The average duration of speech in DiffSSD including real speech is approximately 7.49 seconds (see Table~\ref{tab:dataset-comparison}). 
We split DiffSSD into training, validation and testing sets such that both closed-set and open-set testing scenarios can be analyzed for synthetic speech detection.
In closed-set, the speech generators in the testing set are also present in the training set~\cite{amit_ei_paper}.
In the open-set scenario, synthetic speech from some generators is not used for training of detection methods, which is important in practical situations~\cite{amit_ei_paper}. 
We select five synthetic speech detection methods and demonstrate the significance of DiffSSD in detecting synthetic speech from recent diffusion-based and commercial generators. 
In this paper, we describe the dataset, the input text used for \gls{tts}, and the training, validation and testing splits used in our analysis.

\glsresetall
\section{Related Work}\label{sec:related_works}

Synthetic speech detection methods are broadly divided into three categories~\cite{bhagtani2022overviewrecentworkmedia}. 
Some methods use speech features as input (\eg \gls{lfccs}~\cite{li2021replay} and \gls{mfccs}~\cite{akdeniz2021detection}), others use speech spectrograms or mel-spectrograms~\cite{bartusiak2023}, and the remaining ones use the temporal amplitude of the speech waveform~\cite{tssdnet_2021}.
A spectrogram is a 2D representation of speech as shown in Figure~\ref{fig:spectrogram}~\cite{bartusiak2023}.
It has time on x-axis and frequency in Hertz (Hz) on y-axis~\cite{bartusiak2023}.
If the frequency is in mel (logarithmic) scale, it is called a mel-spectrogram~\cite{mel1937}.
The inputs to synthetic speech detection methods are processed either by \gls{gmm}~\cite{asvdata_2019, lfcc_interspeech}, neural networks such as ResNet~\cite{alzantot2019} or transformer networks~\cite{koutini2022,bartusiak2023}.
The output of these networks is used to determine if the input speech is real or synthetic.

Visual differences between speech generated by two different synthesizers are shown in Figure~\ref{fig:spectrogram}.
The figure shows two spectrograms, one corresponding to a speech from the ASVspoof2019 Dataset~\cite{asvdata_2019} (generated by A11 synthesizer), and the other, to a speech generated using Elevenlabs~\cite{elevenlabs2023}, a commercial software. 
Both speech signals are generated using real speech of the same speaker, and contain spoken words corresponding to the same text. 
As shown in Figure~\ref{fig:spectrogram}, there exist some differences in both spectrograms indicating that speech generators have patterns or artifacts which are unique, and these are leveraged by synthetic speech detection methods. 
Therefore, detectors which are trained to detect synthetic speech from one type of generator may not be able to detect speech from another generator.
In~\cite{bhagtani2024}, the initial study conducted by the authors showed that synthetic speech detection methods trained on conventional speech generators do not generalize to diffusion-based speech generators. 
Only 25,000 synthetic speech signals from 5 synthesizers were used in this analysis~\cite{bhagtani2024}. 
In this paper, we propose and use an extended version of the dataset with 45,000 synthetic speech signals from 5 additional synthesizers. We also re-train the detection methods and examine their performance on DiffSSD. 
\glsresetall

\section{DiffSSD Dataset}\label{sec:dataset}

In this section, we describe the details of the proposed dataset and its development.
\begin{table*}[ht]
    \centering
    \caption{Description of Diffusion-based Synthetic Speech Dataset (DiffSSD). \(\text{S}_{r}\) denotes the sampling rate of speech from each source.}
    \resizebox{0.9\textwidth}{!}{
\renewcommand{\arraystretch}{1.1} 
\begin{tabular}{llrrrccrr}
    \toprule
     & \textbf{Class} & $\boldsymbol{D_{tr}}$ & $\boldsymbol{D_{val}}$ & $\boldsymbol{D_{test}}$ & \textbf{License} & \textbf{Method} &\(\mathbf{S_{r} (kHz)}\) &\textbf{Total}  \\
    \midrule
    \multirow{2}{*}{\makecell{\textbf{Speech} \\\textbf{Signals}}} & \textbf{Real} & 9,690 & 2,423 & 12,113 & & & & \textbf{24,226} \\ 
    & \textbf{Synthetic} & 22,000 & 5,500 & 42,500 & & & & \textbf{70,000} \\
    \midrule
    \multirow{2}{*}{\makecell{\textbf{Speakers}}} & \textbf{Real} & 74 & 74 & 74 & & & & \textbf{74} \\
    & \textbf{Synthetic}  & 5  & 2  & 6 & & & & \textbf{11}\\
    \midrule
    \multirow{12}{*}{\makecell{\textbf{Source}}} 
    & \textbf{LibriSpeech} & \checkmark (4,450) & \checkmark (1,113) & \checkmark (5,563) &  & & 16 & \textbf{11,126} \\
    & \textbf{LJ Speech} & \checkmark (5,240) & \checkmark (1,310) & \checkmark (6,550)  &  & & 22.05 & \textbf{13,100} \\ 
    & \textbf{Elevenlabs} & \checkmark (2,000) & \checkmark (500) & \checkmark (2,500) & CM  & ZS & 44.1 & \textbf{5,000} \\
    & \textbf{GradTTS} & \checkmark (2,000) & \checkmark (500) & \checkmark (2,500) & CM & PT & 22.05 & \textbf{5,000} \\
    & \textbf{Openvoice2} & \checkmark (10,000) & \checkmark (2,500) & \checkmark (12,500) & OS & ZS & 22.05 & \textbf{25,000} \\
    & \textbf{ProDiff} & \checkmark (2,000) & \checkmark (500) & \checkmark (2,500) & OS & PT & 22.05 & \textbf{5,000} \\
    & \textbf{Wavegrad2} & \checkmark (2,000) & \checkmark (500) & \checkmark (2,500) & OS & PT & 22.05 & \textbf{5,000} \\
    & \textbf{Xttsv2 } & \checkmark (2,000) & \checkmark (500) & \checkmark (2,500) & OS & ZS & 24 & \textbf{5,000} \\
    & \textbf{YourTTS} & \checkmark (2,000) & \checkmark (500) & \checkmark (2,500) & OS & ZS & 16 & \textbf{5,000} \\
    & \textbf{DiffGANTTS} & \ding{55} & \ding{55} & \checkmark (5,000) & OS & PT & 22.05 & \textbf{5,000} \\
    & \textbf{PlayHT} & \ding{55} & \ding{55} & \checkmark (5,000) & OS & ZS & 24 & \textbf{5,000} \\
    & \textbf{UnitSpeech} & \ding{55} & \ding{55} & \checkmark (5,000) & OS & ZS & 22.05 & \textbf{5,000} \\
    \midrule
    \textbf{Total} & & \textbf{31,690} & \textbf{7,423} & \textbf{54,613} & & & & \textbf{94,226}\\
    \bottomrule
\vspace{-2em}
\end{tabular}
}

    \label{tab:dataset}
\end{table*}
The description is summarized in Table~\ref{tab:dataset}.
DiffSSD consists of real and synthetic speech divided into training (denoted by $\boldsymbol{D_{tr}}$), validation (denoted by $\boldsymbol{D_{val}}$), and testing sets (denoted by $\boldsymbol{D_{test}}$) as shown in Table~\ref{tab:dataset}. 
Real speech in the dataset is collected from the LJ Speech~\cite{ljspeech17} and LibriSpeech~\cite{ljspeech17} datasets.
All 13,100 speech signals from the LJ Speech dataset are selected.
From the LibriSpeech dataset, speech signals from the development and testing sets are selected, totaling 11,126 speech signals. 

The development of the synthetic speech part of the dataset requires two steps: generation of text input for \gls{tts} methods, and speech generation from text input using the 10 \gls{tts} methods shown in Table~\ref{tab:dataset}, namely Elevenlabs~\cite{elevenlabs2023}, GradTTS~\cite{popov2021gradtts}, Openvoice2~\cite{qin2024openvoice}, ProDiff~\cite{huang2022prodiff}, Wavegrad2~\cite{chen2021wavegrad2}, Xttsv2~\cite{xttsv2}, YourTTS~\cite{casanova2022yourtts}, DiffGANTTS~\cite{Liu2022DiffGANTTSHA}, PlayHT~\cite{playht2023}, and UnitSpeech~\cite{kim2023_unitspeech}.

\subsection{Generation of Text for TTS input}\label{sec:text-gen}
We used ChatGPT 3.5~\cite{radford2018gpt1,ray2023} for generation of 5000 lines of text, with each line containing one or more full sentences in English. 
Topics covered by the generated text (number of lines are in parenthesis) include conversation between people (4275), quotes (169), description of weather (40), animals (68), food (90), news (59), places (76), space (52), sports (96), and history (75). 
Text is processed to avoid repetition.
The lines of text have length varying from 4 to 43 words. 
On average, there are approximately 17 words per line.

\subsection{Synthetic Speech Generation}

Two kinds of \gls{tts} methods are present in DiffSSD: Zero-shot (denoted by ZS in Table~\ref{tab:dataset}) and Pre-trained (denoted by PT in Table~\ref{tab:dataset}).
To generate speech using ZS methods given text, we require only a few minutes of any individual's real speech, and do not need to retrain the \gls{tts} method for that specific individual speaker~\cite{elevenlabs2023,playht2023}.
For PT methods, we require to retrain the methods with hundreds of real speech signals from an individual to be able to generate their high-quality synthetic speech using input text~\cite{huang2022prodiff,Liu2022DiffGANTTSHA}.
All 10 \gls{tts} methods are also categorized based on their license in Table~\ref{tab:dataset}. Two methods are commercial (denoted by CM in Table~\ref{tab:dataset}) tools, which required us to purchase credits for \gls{tts} \gls{vc} during the preparation of this dataset. 
The presence of synthetic speech from commercial methods in DiffSSD is of high significance.
This is because in practical situations \eg for a misinformation campaign on a social media platform, it is easy for attackers with sufficient resources to use these commercial tools for spreading misinformation.
Other \gls{tts} methods in DiffSSD besides commercial software, are open-sourced (denoted by OS in Table~\ref{tab:dataset}), with their source code being publicly available for training.

For generating synthetic speech in DiffSSD using ZS methods, we select 10 speakers (5 female and 5 male) from the LibriSpeech Dataset~\cite{librispeech} and first 500 lines of generated text (described in Section~\ref{sec:text-gen}). 
For each speaker, a few minutes of their real speech from the LibriSpeech Dataset is used and 500 synthetic speech signals are generated with spoken words corresponding to each of the 500 lines of text.
For synthetic speech generation using PT methods, the single speaker in LJ Speech Dataset~\cite{ljspeech17} is selected and for each PT \gls{tts} method (see Table~\ref{tab:dataset}), 5000 synthetic speech signals are generated with spoken words corresponding to each of the 5000 lines of text described in Section~\ref{sec:text-gen}. 
For PT methods, we use weights pre-trained on the LJ Speech Dataset to generate speech.

\begin{table*}[!t]
    \centering
    \caption{Results after training on ASVspoof2019 Dataset, and evaluation on $A_{dev}$, $A_{eval}$, $D_{val}$, and $D_{test}$ sets.} 
    \resizebox{1.0\textwidth}{!}{
        {\renewcommand{\arraystretch}{1.1} 
        \begin{tabular}{l|llrrrr}
	        \toprule
            \textbf{Detection Method} &  \textbf{Input Feature} & \textbf{Processing Network} & $\boldsymbol{A_{dev}}$ & $\boldsymbol{A_{eval}}$ & $\boldsymbol{D_{val}}$ & $\boldsymbol{D_{test}}$ \\
            \midrule
            $\mathrm{LFCC-GMM}$~\cite{asvdata_2019, lfcc_interspeech}  & LFCC & GMM & 0.04\%  & 3.67\% & 22.37\% & 36.73\%\\
            $\mathrm{MFCC-ResNet}$~\cite{alzantot2019} & MFCC & ResNet & 6.52\% & 11.58\% & 52.67\% & 55.06\%\\
            $\mathrm{Spec-ResNet}$~\cite{alzantot2019}  & STFT Magnitude  & ResNet & 0.71\% & 10.10\% & 49.90\% & 52.33\%\\
            $\mathrm{PaSST}$~\cite{bartusiak2023, koutini2022}  & Mel-spectrogram & Transformer network & 4.10\% & 5.26\% & 35.99\% & 32.25\% \\
            $\mathrm{Wav2Vec2}$~\cite{tak22_odyssey}  & Temporal amplitude  & Transformer network & 0.02\% & 0.30\% & 46.18\% & 48.53\% \\
	    \bottomrule
        \end{tabular}}}
    \label{tab:exp-asv19}
\vspace{-2em}
\end{table*}
\subsection{Creating Training, Validation, and Testing Sets}
As described in Section~\ref{sec:intro}, we create $\boldsymbol{D_{tr}}$, $\boldsymbol{D_{val}}$, and $\boldsymbol{D_{test}}$ sets for training, validation, and testing, respectively such that closed-set and open-set scenarios can be analyzed for synthetic speech detection.
All real speech signals in DiffSSD are randomly divided into $\boldsymbol{D_{tr}}$, $\boldsymbol{D_{val}}$, and $\boldsymbol{D_{test}}$ in the ratio 40:10:50, respectively.
In the case of synthetic speech, 7 out of 10 \gls{tts} methods are present in all three sets for closed-set analysis. 
3 \gls{tts} methods, which also include a commercial software, and not used for training and validation, but are used for testing. 
This is useful for open-set analysis.
For ZS methods, 4 speakers are used for training, 1 for validation, and 5 are used for testing. This ensures that speakers do not overlap among the 3 sets. 
For PT methods, since there is only a single speaker, all speech signals are randomly divided into $\boldsymbol{D_{tr}}$, $\boldsymbol{D_{val}}$, and $\boldsymbol{D_{test}}$ in the ratio 40:10:50, respectively.

\glsresetall

\begin{table}[!b]
    \centering
    \vspace{-2em}
    \caption{Results after training on $D_{tr}$ set of DiffSSD.}
    \resizebox{0.85\columnwidth}{!}{
        {\renewcommand{\arraystretch}{1.1} 
        \begin{tabular}{l|rr}
	        \toprule
            \textbf{Detection Method} & $\boldsymbol{D_{val}}$ & $\boldsymbol{D_{test}}$ \\
            \midrule
            $\mathrm{LFCC-GMM}$ & 25.20\%  & 22.04\% \\
            $\mathrm{MFCC-ResNet}$ & 4.37\% & 6.69\% \\
            $\mathrm{Spec-ResNet}$  & 1.69\% & 11.00\% \\
            $\mathrm{PaSST}$  & 0.08\% & 3.53\% \\
            $\mathrm{Wav2Vec2}$  & 1.51\% & 3.00\%  \\
	    \bottomrule
        \end{tabular}}}
    \label{tab:exp-diffssd}
\vspace{-1em}
\end{table}

\section{Experiments and Results}\label{sec:experiments}
In this section, we examine the performance of synthetic speech detectors on the ASVspoof2019 Dataset and DiffSSD.
\subsection{Evaluation Metric}
We use \gls{eer}, the primary evaluation metric in the ASVspoof2019 Challenge~\cite{asvdata_2019}.
\gls{eer} is defined as the False Acceptance Rate (FAR) on the Receiver Operating Characteristics (ROC) curve where FAR is equal to False Rejection Rate (FRR).
Lower the EER, the better is the performance of the method.

\subsection{Synthetic Speech Detection}
We select 5 detection methods which have shown high performance on the ASVspoof2019 Dataset and are representative of most categories described in Section~\ref{sec:related_works}. 
We select 2 methods which use speech features \gls{lfccs}~\cite{asvdata_2019, lfcc_interspeech} and \gls{mfccs}~\cite{alzantot2019}, two methods, each of which process mel-spectrogram~\cite{bartusiak2023} and magnitude of \gls{stft}~\cite{alzantot2019} and one which processes temporal amplitude of the speech waveform~\cite{tak22_odyssey}. 
All detection methods used in our analysis are shown in Table~\ref{tab:exp-asv19}.
We trained all these detection methods on the training set of ASVspoof2019, and evaluated them on the development (denoted by $\boldsymbol{A_{dev}}$) and evaluation (denoted by $\boldsymbol{A_{eval}}$) sets of the ASVspoof2019 dataset~\cite{asvdata_2019} as shown in Table~\ref{tab:exp-asv19}.
We also evaluated them on the validation ($\boldsymbol{D_{val}}$) and testing ($\boldsymbol{D_{test}}$) sets of DiffSSD.
The results of this analysis are shown in Table~\ref{tab:exp-asv19}. 
We observe that all detection methods with near-perfect performance on the ASVspoof2019 dataset show a decline in performance when evaluated on DiffSSD indicating poor generalization performance for diffusion-based and commercial generators.

\begin{figure}[h]
\begin{center}
    \vspace{-1em}
    \includegraphics[width=\columnwidth]{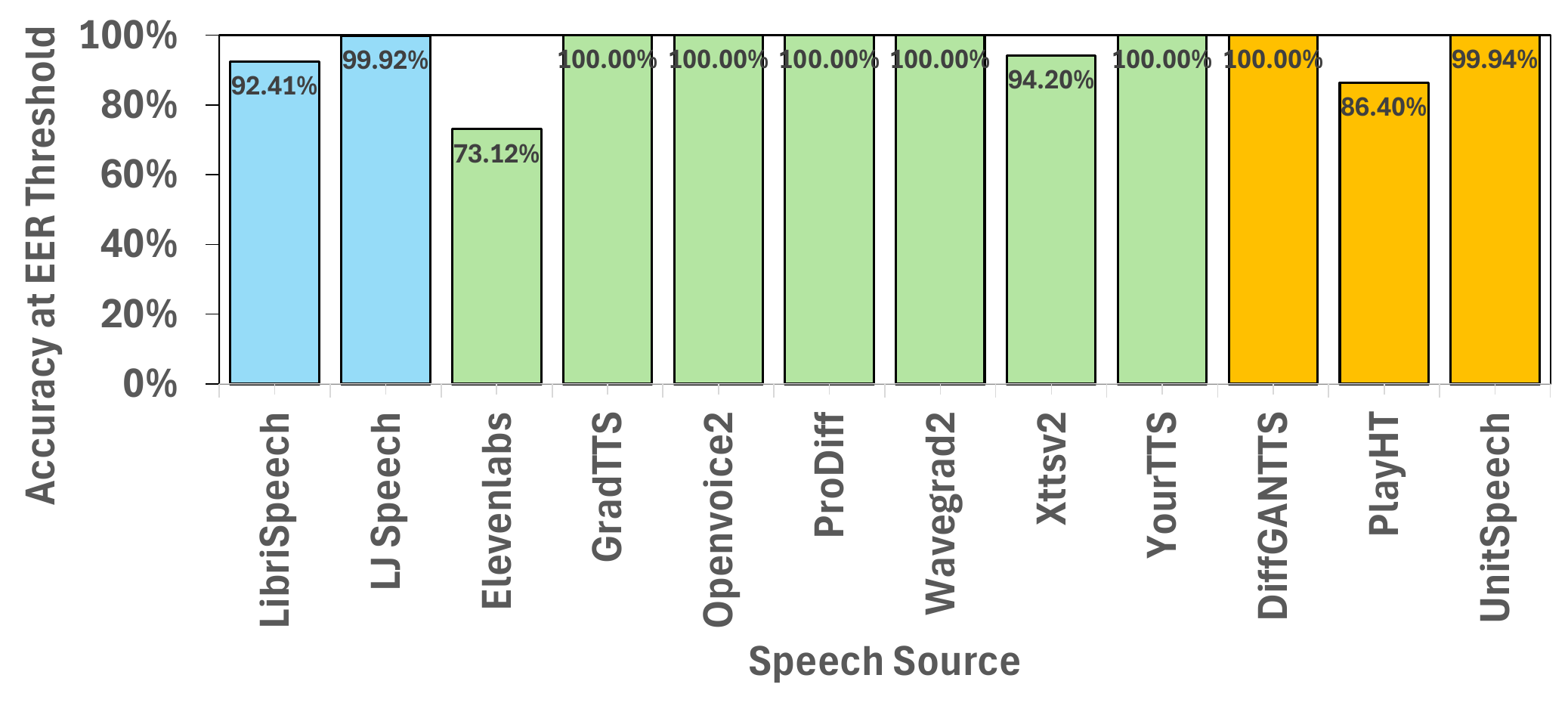}
    \vspace{-2em}
\caption{PaSST performance for each synthesizer in DiffSSD.}
\label{fig:individual-performance}
\end{center}
\vspace{-2em}
\end{figure}

In our next analysis, we re-train these methods on the $\boldsymbol{D_{tr}}$ set and re-evaluate on $\boldsymbol{D_{val}}$ and $\boldsymbol{D_{test}}$. The results of this analysis are shown in Table~\ref{tab:exp-diffssd}.
Except LFCC-GMM, all detection methods demonstrate significant improvement in performance for detecting synthetic speech from recent synthesizers. 
One of the reasons behind poor performance of LFCC-GMM could be the presence of high-quality synthetic speech in DiffSSD, which is perceptually indistinguishable from real speech.
Perhaps handcrafted LFCC features are not sufficient to capture the differences between both categories. 
Besides cumulative performance as shown in Table~\ref{tab:exp-diffssd}, we also evaluate the performance of one of the detectors, PaSST w.r.t each \gls{tts} synthesizer present in the dataset as shown in Figure~\ref{fig:individual-performance}.
We use Accuracy at EER Threshold to measure performance in Figure~\ref{fig:individual-performance}.
This refers to the detection method's accuracy at the~\gls{eer} decision threshold.
In Figure~\ref{fig:individual-performance}, blue represents real speech, green represents the detectors present during training (closed-set scenario), and yellow represents the detectors not used during training (open-set scenario). 
We observe that the method shows perfect detection for most generators including the ones not used in training. 
With this analysis, we show the significance of our proposed dataset in detecting synthetic speech generated from
recent diffusion-based generators and commercial software.

\glsresetall

\section{Conclusion and Future Work}\label{sec:conclusion}

In this paper, we proposed the Diffusion-Based Synthetic Speech Dataset (DiffSSD). We examined synthetic speech detectors using this dataset. Future work should focus on developing detectors which can detect high-quality synthetic speech from advanced commercial tools, such as Elevenlabs, and show even better performance in the open-set scenario.

The link to DiffSSD:\url{https://huggingface.co/datasets/purdueviperlab/diffssd}.
We are distributing the speech data, the text used as input to generate data, and the training, validation, and testing splits.

\section*{Acknowledgments}
This material is based on research sponsored by DARPA
and Air Force Research Laboratory (AFRL) under agreement
number FA8750-20-2-1004. The U.S. Government is authorized to reproduce and distribute reprints for Governmental
purposes notwithstanding any copyright notation thereon.
The views and conclusions contained herein are those of the
authors and should not be interpreted as necessarily representing the official policies or endorsements, either expressed or implied, of DARPA and Air Force Research Laboratory (AFRL) or the U.S. Government. Address all correspondence to Edward J. Delp, ace@purdue.edu.

\section*{References}
\setlength{\bibitemsep}{0.3ex}
\AtNextBibliography{\fontsize{7.5}{9}\selectfont}
\printbibliography[heading=none]

\end{document}